\documentclass[preprint,aps,tightenlines,showpacs,superscriptaddress]{revtex4}
\usepackage[dvips]{graphicx}
\usepackage{dcolumn}
\usepackage{epsfig}

\begin{document}
\title{Extraction of $P_{11}$ resonances from $\pi N$ data}
\author{H. Kamano}
\affiliation{Excited Baryon Analysis Center (EBAC), Thomas Jefferson National
Accelerator Facility, Newport News, Virginia 23606, USA}
\author{S. X. Nakamura}
\affiliation{Excited Baryon Analysis Center (EBAC), Thomas Jefferson National
Accelerator Facility, Newport News, Virginia 23606, USA}
\author{T.-S. H. Lee}
\affiliation{Excited Baryon Analysis Center (EBAC), Thomas Jefferson National
Accelerator Facility, Newport News, Virginia 23606, USA}
\affiliation{Physics Division, Argonne National Laboratory,
Argonne, Illinois 60439, USA}
\author{T. Sato}
\affiliation{Department of Physics, Osaka University, Toyonaka,
Osaka 560-0043, Japan}
\affiliation{Excited Baryon Analysis Center (EBAC), Thomas Jefferson National
Accelerator Facility, Newport News, Virginia 23606, USA}

\begin{abstract}
We show that two $P_{11}$ nucleon resonance poles near the $\pi \Delta$
threshold, obtained in several analyses,
are stable against large variations of parameters within a 
dynamical coupled-channels analysis based on meson-exchange mechanisms. 
By also performing an analysis based on a model with a bare nucleon state,
we find that this two-pole structure is insensitive to the analytic structure
of the amplitude in the region below $\pi N$ threshold.
Our results are $M_{\text{pole}} = (1363^{+9}_{-6}\, -i\, 79^{+3}_{-5})$ MeV 
and $(1373^{+12}_{-10}\, -i\, 114^{+14}_{-9})$ MeV. 
We also demonstrate that the number of  poles in the
1.5 GeV $\leq W \leq $ 2 GeV region could be more than one, 
depending on how the structure of the single-energy solution of SAID 
is fitted. For three-pole solutions, our best estimated result
of a pole near $N(1710)$ listed by Particle Data Group is
$(1829^{+131}_{-65}\, -i\, 192^{+88}_{-110})$ MeV which is close to
the results of several previous analyses.
Our results indicate the need of more accurate $\pi N$ reaction
data in the $W > $ 1.6 GeV region for high precision resonance extractions. 
\end{abstract}

\maketitle

\section{Introduction}
\label{sec1}

An important task in hadron physics is to extract
nucleon resonances from $\pi N$ reaction data. The extracted resonance
parameters are needed to understand the spectrum and 
structure of excited nucleons within QCD. They are also the starting 
point for analyzing electromagnetic meson production reaction data which
have been of high precision and extensive in recent years~\cite{bl04}.

 There exists several 
approaches~\cite{cmb79,cw90,kh83,said-0,said-1,ksu,zegrab-1,zegrab-2,pitt-anl,ntu-mainz,bonn,juelich,sjklms10} to extract nucleon resonances ($N^*$)
from $\pi N$ reaction data. In general,  almost all
4-stars nucleon resonances  listed by Particle Data Group~\cite{pdg} (PDG)
are found
in all approaches. However the existence of some $N^*$ states, in particular
those in the higher mass region, is controversial. The most investigated
case is the number of  resonances in $\pi N$ $P_{11}$ partial wave.
In the region near Roper $N(1440)$, two poles
close to the $\pi\Delta$ threshold were found in 
Refs.~\cite{said-0,cw90,said-1,juelich}
and in our recent extraction~\cite{sjklms10}, while only one pole in the
similar energy region was reported in
Refs.~\cite{kh83,zegrab-2,pitt-anl}.
In the higher mass region, the $N(1710)$ in $P_{11}$ $\pi N $ partial-wave
 is not reported in
Refs.~\cite{said-1,juelich}, but is identified in all other
analysis~\cite{kh83,cmb79,cw90,ksu,zegrab-1,zegrab-2,pitt-anl,ntu-mainz,sjklms10}.
 
To make progress, it is important to address a commonly asked question on
the extent to which the extracted resonance parameters depend on
the reaction models employed and the accuracy of the empirical partial-wave
amplitudes used in the analysis. For $P_{11}$ resonances, this was
investigated by Cutkosky and Wang~\cite{cw90} and
more recently  by Ceci, Svarc, and Zauner~\cite{zegrab-2} within the 
Carnegie-Mellon University-Berkeley model~\cite{cmb79} (CMB). 
In an
analysis including $\pi N, \eta N$ and pseudo $\pi\pi N$ channels, it was
demonstrated~\cite{zegrab-2} that
the existence of $N(1710)$ depends on the structure of the $\pi N$
amplitude which is related to the coupled-channels effects due to
$\eta N$ channel.
In this work, we carry out a similar investigation within a dynamical
coupled-channels model~\cite{msl07} (EBAC-DCC). 
The main difference between our
approach and CMB model is to define the non-resonant amplitude by using
the meson-exchange mechanisms. 
We thus have provided additional
information for examining the dependence of the $P_{11}$ resonances on 
the reaction models employed in the analysis.

Our investigation has two parts. 
First we would like to examine the stability
of the two-pole structure of $P_{11}$ resonances
 near the $\pi \Delta$ threshold ($W \sim 1.3$ GeV).
Our objective is to examine how this two-pole structure 
is sensitive to
the parameters of the meson-exchange mechanisms within
the EBAC-DCC model  used in our
extraction~\cite{sjklms10}.
In the fits~\cite{jlms07} (JLMS) of $\pi N$ data~\cite{said-1},
these parameters were determined within the ranges 
known from previous studies of meson-exchange mechanisms.
Here we allow them to vary much more freely such that several models
with different analytic properties are obtained for examining whether 
the resulting pole positions are stable within the EBAC-DCC model.

The two-pole structure is also reported by 
D{\"o}ring {\it et al.}~\cite{juelich} in an analysis based on  a 
meson-exchange model with a bare nucleon state~\cite{juelich-1}.
As discussed in Ref.~\cite{jklmss09},  the analytic structure
of this model as well as other similar models~\cite{afnan,gross,ntuanl} 
is  rather different from the EBAC-DCC model, in particular in the region near
the nucleon pole, mainly because of the differences in deriving~\cite{kleinlee}
three-dimensional scattering equations from relativistic quantum field theory.
To further examine the stability
of the two-pole structure of $P_{11}$ resonances and the existence of
$N(1710)$ state within 
the meson-exchange models,
we also perform fits by using such a  model. Our formulation is similar to
that developed by Pearce and Afnan~\cite{afnan}.

We will show that 
the positions of two poles near the $\pi \Delta$
threshold extracted from all of the meson-exchange models constructed here
are rather stable.
This explains why the similar two-pole structure
is also found in the other analyses~\cite{cw90,said-1,juelich}
 which use very different reaction models.

The second part of our investigation is 
to examine the extent to which the structure of the $P_{11}$ amplitude
in higher invariant mass $(W)$ region can influence the two-pole 
structure near the $\pi\Delta$ threshold.
Here we also follow Ref.~\cite{zegrab-2} to 
examine how the number of
resonance states in the region near $N(1710)$ state listed by PDG
depend on the  structure of the data.
We thus consider both the energy-dependent
and single-energy solutions (SP06) of SAID~\cite{said-1},
hereafter we call them SAID-EDS and SAID-SES, respectively,
as well as a solution from CMB~\cite{cw90} collaboration. 
The CMB amplitudes could be outdated, but
is used here only for investigating the dependence of the $P_{11}$ poles on the
accuracy of the data.
We will show that the number of
resonance poles in the 1.6 GeV $< W <$ 2 GeV could be more than one,
depending on how the structure of the amplitude
is fitted. Our results indicate the importance of improving the accuracy
of empirical partial-wave amplitudes. More accurate $\pi N$ reaction data from
the new hadron facilities, such as the Japan Proton Accelerator Research Complex (J-PARC), 
are needed.
Our conclusion is consistent with the finding of Ref.~\cite{zegrab-2} in 
which the importance of also fitting the
$\pi N \rightarrow \eta N$ amplitude is demonstrated in identifying $P_{11}$
$N(1710)$ state.

In Sec.~\ref{sec2}, we give a brief description of the
coupled-channels models used in this work. 
The results are given and discussed in Sec.~\ref{sec3}. 
Section~\ref{sec4} is devoted to the discussions 
on possible further developments.

\section{Dynamical Coupled-channels models}
\label{sec2}

In this section we first recall briefly   the
EBAC-DCC model~\cite{msl07} used in this work.
We then describe how the model can be modified to obtain 
a model with  a bare nucleon, which has the main feature
of other $\pi N$ reaction models with a bare 
nucleon~\cite{afnan,gross,juelich,ntuanl}.

\subsection{EBAC-DCC model}
\label{sec2a}

The EBAC-DCC model describes meson-baryon reactions 
involving the following channels: $\pi N$, $\eta N$, and $\pi\pi N$ which
has $\pi\Delta$, $\rho N$, and $\sigma N$ resonant components. The
excitation of the internal structure of a baryon ($B$) by a meson
($M$) to a bare $N^*$ state is modeled by a vertex interaction
$\Gamma_{MB \leftrightarrow N^*}$. The meson-baryon ($MB$) states can interact
via interactions $v_{MB,M'B'}$ which describe the meson-exchange
mechanisms deduced from phenomenological Lagrangians.
Within the model,  the partial-wave amplitude
of the $M(\vec k)+ B(-\vec k) \to M'(\vec k')+ B'(-\vec k')$ reaction
can be cast into the 
following form (suppressing the angular momentum and isospin
indices):
\begin{eqnarray}
T_{MB,M'B'}(k,k',E)  &=&  t_{MB,M'B'}(k,k',E) + t^{R}_{MB,M'B'}(k,k',E),
\label{eq:tmbmb}
\end{eqnarray}
where the first term is defined by a set of coupled-channels integral
equations
\begin{eqnarray}
t_{MB,M^\prime B^\prime}(k,k',E) &=&  v_{MB,M^\prime B^\prime}(k,k')
\nonumber\\
&+& \sum_{M^{\prime\prime}B^{\prime\prime}}
\int_{C_{M^{\prime\prime}B^{\prime\prime}}} q^2 dq
v_{MB,M^{\prime\prime}B^{\prime\prime}}(k,q)
G_{M^{\prime\prime}B^{\prime\prime}}(q,E)
t_{M^{\prime\prime}B^{\prime\prime},M^\prime B^\prime}(q,k',E).
\label{eq:cc-mbmb}
\end{eqnarray}
Here $C_{MB}$ is the integration contour in the complex-$q$ plane used
for the channel $MB$.
The term associated with the bare $N^*$ states in Eq.~(\ref{eq:tmbmb}) is
\begin{eqnarray}
t^{R}_{MB,M^\prime B^\prime}(k,k',E)&=& \sum_{i,j}
\bar{\Gamma}_{MB \to N^*_i}(k,E) [D(E)]_{i,j}
\bar{\Gamma}_{N^*_j \to M^\prime B^\prime}(k',E),
\label{eq:tmbmb-r}
\end{eqnarray}
where the dressed vertex function 
$\bar{\Gamma}_{N^*_j \to M^\prime B^\prime}(k,E)$ is 
 calculated~\cite{jlms07} from
the bare vertex ${\Gamma}_{N^*_j \to M^\prime B^\prime}(k)$ and convolutions
over the amplitudes $t_{MB,M^\prime B^\prime}(k,k',E)$.
The inverse of the propagator of dressed $N^*$ states in
Eq.~(\ref{eq:tmbmb-r})
is \begin{equation}
[D^{-1}(E)]_{i,j} = (E - m^0_{N^*_i})\delta_{i,j} - \Sigma_{i,j}(E) ,
\label{eq:nstar-selfe}
\end{equation}
where $m^0_{N^*_i}$  is the bare mass of the $i$-th $N^*$ state,
and the $N^*$ self-energy is defined by
\begin{equation}
\Sigma_{i,j}(E)= \sum_{MB} \int_{C_{MB}}  q^2 dq 
\bar{\Gamma}_{N^*_j \to M B}(q,E) G_{MB}(q,E) {\Gamma}_{MB \to N^*_i}(q,E).
\label{eq:nstar-g}
\end{equation}
Defining  $E_\alpha(k)=[m^2_\alpha + k^2]^{1/2}$ with $m_\alpha$ being
the mass of particle $\alpha$,
the meson-baryon propagators in the above equations are:
$G_{MB}(k,E)=1/[E-E_M(k)-E_B(k) + i\epsilon]$ for the stable
$\pi N$ and $\eta N$ channels, and $G_{MB}(k,E)=1/[E-E_M(k)-E_B(k) -\Sigma_{MB}(k,E)]$
for the unstable $\pi\Delta$, $\rho N$, and $\sigma N$ channels.
The self energy $\Sigma_{MB}(k,E)$ is calculated from a vertex
function defining the decay of the considered unstable particle
in the presence of a spectator $\pi$ or $N$ with momentum $k$.
For example, we have for the $\pi\Delta$ state,
\begin{eqnarray}
\Sigma_{\pi\Delta}(k,E) &=&\frac{m_\Delta}{E_\Delta(k)}
\int_{C_3} q^2 dq \frac{ M_{\pi N}(q)}{[M^2_{\pi N}(q) + k^2]^{1/2}}
\frac{\left|f_{\Delta \to \pi N}(q)\right|^2}
{E-E_\pi(k) -[M^2_{\pi N}(q) + k^2]^{1/2} + i\epsilon},
\label{eq:self-pid}
\end{eqnarray}
where $M_{\pi N}(q) =E_\pi(q)+E_N(q)$ and $f_{\Delta \to \pi N}(q)$
defines the decay of the $\Delta \to \pi N$ in the rest frame
of $\Delta$, $C_3$ is the corresponding integration contour in the
complex-$q$ plane. The self-energies for the $\rho N$ and $\sigma N$ channels
are similar.

To search for resonance poles,  the contours
$C_{MB}$ and $C_3$ must be chosen appropriately 
to solve Eqs.~(\ref{eq:cc-mbmb})-(\ref{eq:self-pid})
for $E$ on the various possible sheets of the Riemann surface.
The procedures for performing this numerical task have been discussed
in Ref.~\cite{ssl09,sjklms10}.
Like all previous works~\cite{said-1,pitt-anl}, we only look for poles
which are close to the physical region and have effects on the $\pi N$
scattering  observables. All of these poles are on the unphysical sheet
of the $\pi N$ channel, but could be on either unphysical $(u)$ or
physical $(p)$ sheets of other channels considered in this analysis.
We will indicate the sheets where the identified poles are located by
$(s_{\pi N}, s_{\eta N}, s_{\pi \pi N} ,s_{\pi \Delta},s_{\rho N},
s_{\sigma N})$, where $s_{MB}$ and $s_{\pi\pi N}$ can be
$u$ or $p$.

\subsection{Model with a bare nucleon state}
\label{sec2b}

To examine further the model dependence of resonance extractions, 
it is useful to also  perform
analysis using models with a bare nucleon, as developed in, for example,
Refs.~\cite{afnan,gross,ntuanl,juelich}.
Within the formulation given in Sec.~\ref{sec2a},
such a model can be obtained by
adding a bare nucleon ($N_0$) state with mass $m^0_N$
and $N_0\rightarrow MB $ vertices and
removing the direct $MB \rightarrow N \rightarrow M'B'$
in the meson-baryon
interactions $v_{MB,M'B'}$.
All numerical procedures for this model
are identical to that used in the JLMS analysis,
except that the resulting amplitude must satisfy the nucleon pole condition.
Here we follow the procedure of Afnan and Pearce~\cite{afnan}.

For simplicity, we include one bare $N_0$ state and only one bare $N^*$ state.
The amplitude can still be written in the form of Eq.~(\ref{eq:tmbmb})
and the propagator $D(E)$ of the term  $t^R$ of Eq.~(\ref{eq:tmbmb-r})
is a $2\times 2$ matrix.
The nucleon pole condition can be most transparently defined by introducing an
orthogonal matrix ${U}^TU=1$ ($U^T_{ij}= U_{ji}$) to diagonalize 
$D^{-1}(E)$ of Eq.~(\ref{eq:nstar-selfe}). 
The term $t^{R}_{\pi N,\pi N}$ of Eq.~(\ref{eq:tmbmb-r}) can then be
cast into the following diagonal form
\begin{eqnarray}
t^{R}_{\pi N,\pi N}(k,k,E) =
\sum_{i=1,2} \tilde{t}^i_{\pi N,\pi N}(k,k,E),
\label{eq:tR-diag}
\end{eqnarray}
with
\begin{equation}
\tilde{t}^i_{\pi N,\pi N}(k,k,E)  = \frac{\tilde{F}_{\pi N,i}(k) 
\tilde{F}_{i,\pi N}(k)}
{E-m^0_i-\tilde{\Sigma}_i(E)} ,
\label{eq:diagr}
\end{equation}
where  $m^0_1 = m^0_N$ and $m^0_2 = m^0_{N^*}$, and the mass shifts are
\begin{eqnarray}
\tilde{\Sigma}_1(E) &=& \frac{1}{2}
\{m^0_2-m^0_1+\Sigma_{11}(E) + \Sigma_{22}(E)  \nonumber \\
& &- [(m^0_2+\Sigma_{22}(E) - m^0_1 -\Sigma_{11}(E))^2
+4\Sigma_{12}^2(E)]^{1/2}\}\,,
\label{eq:tilde-sig1},
\\
\tilde{\Sigma}_2 (E) &=& \frac{1}{2}
\{m^0_1-m^0_2+\Sigma_{11}(E) + \Sigma_{22}(E)  \nonumber \\
& &+ [(m^0_2+\Sigma_{22}(E) - m^0_1 -\Sigma_{11}(E))^2
+4\Sigma_{12}^2(E)]^{1/2}\}.
\label{eq:tilde-sig2}
\end{eqnarray}
Here $\Sigma_{i,j}(E)$ are defined by Eq.~(\ref{eq:nstar-g}).
The transformed vertices in Eq.~(\ref{eq:diagr}) are
\begin{eqnarray}
\tilde{F}_{i,\pi N}(k)&=& \sum_{j} U_{i,j} \bar{\Gamma}_{N^*_j \to\pi N}(k) \,,\\
\tilde{F}_{\pi N,i}(k)&=&\sum_{j} U_{i,j}  \bar{\Gamma}_{\pi N\to N^*_j}(k),
\end{eqnarray}
where $N^*_j = N_0$ or $N^*$, and the transformation operator $U$ is defined by
\begin{eqnarray}
U_{11}&=&U_{22}=\frac{1}{(1+\nu^2)^{1/2}} ,\\
U_{12}&=&-U_{21} = \frac{\nu}{(1+\nu^2)^{1/2}},
\end{eqnarray}
with
\begin{equation}
\nu= \frac{\Sigma_{11}-\tilde{\Sigma}_1}{\Sigma_{12} }
=-\frac{\Sigma_{22}-\tilde{\Sigma}_2}{\Sigma_{12}}.
\end{equation}

Suppose $E=m_N$ pole is found in the first term of Eq.~(\ref{eq:tR-diag}),
we then expand
\begin{eqnarray}
E-m^0_1-\tilde{\Sigma}_1(E) &=& E-m^0_1-
\left\{
\tilde{\Sigma}_1(m_N)+
\left[\frac{\partial}
{\partial E} \tilde{\Sigma}_1(E)
\right]_{E=m_N}(E-m_N) 
+ \cdots\right\}
\nonumber \\
&=& (E-m_N) 
\left\{1 -
\left[\frac{\partial}
{\partial E} \tilde{\Sigma}_1(E)
\right]_{E=m_N} + \cdots
\right\} \,,
\end{eqnarray}
where we have defined the nucleon pole
\begin{eqnarray}
m_N=m^0_1 + \tilde{\Sigma}_1(m_N) \,.
\label{eq:pole-m}
\end{eqnarray}
This is the first nucleon pole condition taken into account
in constructing the bare nucleon model.

Defining the renormalized vertex as
\begin{eqnarray}
F_{\pi NN} (k)&=& \tilde{F}_{1,\pi N}(k)Z^{-1/2} \nonumber \\
&=&\sum_{j} U_{1,j} \bar{\Gamma}_{j,\pi N}(k)Z^{-1/2}, 
\end{eqnarray}
with
\begin{equation}
Z= 1 - \left[\frac{\partial} {\partial E} \tilde{\Sigma}_1(E)\right]_{E=m_N},
 \end{equation}
we then have
\begin{equation}
\tilde{t}^R_i(k\to k_{\text{on}},k\to k_{\text{on}},E\rightarrow m_N ) 
= -\frac{[F_{\pi NN} (k_{\text{on}})]^2}{E-m_N} .
\label{eq:pole-t}
\end{equation}
Here the on-shell momentum is defined by 
$E=\sqrt{m_N^2+k^2_{\text{on}}}+\sqrt{m_\pi^2+k^2_{\text{on}}}$.
Below $E=m_N+m_\pi$, $k_{\text{on}}$ becomes positive or negative imaginary.
Here we take the positive imaginary 
since we look for the physical nucleon pole.
The second nucleon pole condition then defines the renormalized vertex
$F_{\pi NN} (k_{\text{on}})$
as the physical $\pi NN$ form factor. Following the partial-wave
decomposition procedure given in Ref.~\cite{msl07}, we find
\begin{equation}
F_{\pi NN}(k_{\text{on}})= F^{\text{phys.}}_{\pi NN}(k_{\text{on}}),
\label{eq:pinn-ff}
\end{equation}
with
\begin{equation}
F^{\text{phys.}}_{\pi NN}(k)=
-\frac{i}{(2\pi)^{3/2}}\frac{f_{\pi NN}}{m_\pi}
\sqrt{12\pi}k\sqrt{E_N(k)+m_N\over 2 E_N(k)}
{1\over \sqrt{2\omega_\pi(k)}} 
\left[1+{E_\pi(k)\over E_N(k)+m_N}\right],
\label{eq:pinn-ff2}
\end{equation}
where $f_{\pi NN}=\sqrt{4\pi\times 0.08}$.
Following the previous approach, the bare $N_0\rightarrow \pi N$ vertex
$\Gamma_{N_0,\pi N}(k)$ is parameterized as Eq.~(\ref{eq:pinn-ff2}) 
except that $f_{\pi NN}$ is replaced by a bare
coupling constant $f^0_{\pi NN}$, and the form factor is introduced.
Explicitly, it is written as
\begin{equation}
\Gamma_{N_0,\pi N}(k) =
-\frac{i}{(2\pi)^{3/2}}\frac{f^0_{\pi NN}}{m_\pi}
\sqrt{12\pi}k\sqrt{E_N(k)+m_N\over 2 E_N(k)}
{1\over \sqrt{2\omega_\pi(k)}} 
\left[1+{E_\pi(k)\over E_N(k)+m_N}\right]
F(k,\Lambda_{\pi NN}),
\label{bare-N-form}
\end{equation}
where we use the following form factor,
\begin{equation}
F(k,\Lambda_{\pi NN}) = 
\left(\frac{\Lambda^2_{\pi NN}}{k^2+\Lambda^2_{\pi NN}}\right)^2.
\end{equation}
The cutoff parameter $\Lambda_{\pi NN}$ of the form factor and 
the bare coupling constant $f^0_{\pi NN}$ are varied along with other parameters
of the model to fit the empirical $\pi N$ scattering amplitudes and the pole 
conditions~(\ref{eq:pole-m}) and~(\ref{eq:pinn-ff}).

Here we note that the pole condition~(\ref{eq:pole-m}) depends on
both $m^0_1$ and $m^0_2$ as can be seen in Eq.~(\ref{eq:tilde-sig1}) for
$\tilde{\Sigma}_1(E)$. Thus the mass renormalization of the physical
nucleon includes not only the meson cloud
effects, but also the contribution from the bare $N^*$ state.
If we drop the $N^*$ state,
the nucleon pole condition become the usual form
\begin{equation}
m_N = m^0_1 +\Sigma_{11}(m_N).
\label{eq:pol-m1}
\end{equation}
We use the exact conditions~(\ref{eq:pole-m}) and~(\ref{eq:pinn-ff}) 
in our investigations. Our approach is not completely consistent with
the rigorous approach of Ref.~\cite{afnan}, but is sufficient for our
present limited purpose of investigating model dependence of resonance 
extractions. Qualitatively, this model contains the main feature of
 the coupled-channels model developed in Ref.~\cite{juelich} in handling
the $\pi N$ scattering in $P_{11}$ partial wave. The main difference is in the 
derivation of meson-baryon potential $v_{MB,M'B'}$ from phenomenological
Lagrangians, as discussed in Ref.~\cite{jklmss09}.

\section{Result}
\label{sec3}

\begin{table}[t]
\caption{\label{tab:p11-tab1}
The resonance pole positions $M_R$ for $P_{11}$
[listed as ($\text{Re}M_R$, $-\text{Im} M_R$) in the unit of MeV] extracted from
 various parameter sets.
The location of the pole is specified by, e.g.,
$(s_{\pi N},s_{\eta N},s_{\pi\pi N},s_{\pi\Delta},s_{\rho N},s_{\sigma N})=(upuupp)$,
where $p$ and $u$ denote the physical and unphysical sheets for a
given reaction channel, respectively. $\chi^2_{pd}$ is
defined by Eq.~(\ref{eq:chi2}).  }
\begin{ruledtabular}
\begin{tabular}{ccccccc}
Model           & $upuupp$   & $upuppp$  & $uuuupp$  & $uuuuup$& $\chi^2_{pd}$   \\ 
\hline
SAID-EDS        & (1359, 81) & (1388, 83) &    ---    &  ---        & 2.94 \\
JLMS            & (1357, 76) & (1364, 105)&    ---    & (1820, 248) & 3.55 \\
1$N^*$-3p-H     & (1357, 74) & (1363, 111)&    ---    & (1792, 280) & 2.41 \\
1$N^*$-3p-L     & (1359, 69) & (1371, 112)&    ---    & (1940, 242) & 5.33 \\
\hline
2$N^*$-3p       & (1368, 82) & (1375, 110)&    ---    & (1810, 82)  & 3.28 \\
2$N^*$-4p       & (1372, 80) & (1385, 114)& (1636, 67)& (1960, 215) & 3.36 \\
\hline
2$N^*$-4p-CMB   & (1379, 89) & (1386, 109)& (1613, 42)& (1913, 324) & 4.91 \\
\hline
1$N_0$1$N^*$-3p & (1363, 81) & (1377, 128)&    ---    & (1764, 137) & 2.51 \\
\end{tabular}
\end{ruledtabular}
\end{table}

 We first discuss the parameters of the coupled-channels
models described in section II, which are varied in performing
$\chi^2$ fits to empirical $P_{11}$ amplitudes using MINUIT.
The  non-resonant amplitude $t_{MB,M'B'}$ of Eq.~(\ref{eq:tmbmb})
is determined by the coupling constants and cutoffs of
form factors of the 
meson-exchange interactions $v_{MB,M'B'}$ through solving the
coupled-channels integral equation (\ref{eq:cc-mbmb}).
In the JLMS fit~\cite{jlms07} to the $\pi N$ data,
these parameters were constrained within the ranges
known from previous studies of meson-exchange mechanisms, as discussed in 
Ref.~\cite{msl07}.
Here we allow them to vary much more freely such that several models
are obtained for examining whether
the resulting pole positions are stable against the variation of
the  analytic properties of the resulting amplitudes.

In the absence of theoretical input, our main challenge 
is to determine the bare $N^*$ mass $m^0_{N^*}$ and
the $N^* \rightarrow MB$ vertex function. 
For $P_{11}$ partial wave, 
the number of $N^*$ parameters is : 
$\displaystyle N_{N^*} +N_{N^*}\times\sum_{MB} n_{v,MB}$,
where $N_{N^*}$ is the number of the bare $N^\ast$
and $n_{v,MB}$ is the number of parameters needed to parameterize
each $N^*\rightarrow MB$ vertex function $\Gamma_{N^*\rightarrow MB}$. 
In our fit we have
$N_{N^*}=1$ or 2, and 
$n_{v,MB}=2$ ($MB=\pi N,\eta N,\pi\Delta,\sigma N$) or 4 ($MB=\rho N$) 
from the coupling constants $g_{MB}$ and cutoffs $\Lambda_{MB}$
(as explained in Ref.~\cite{jlms07}).
We have total five channels ($N_{MB}=5$).
We thus face a many-parameters problem in fitting the data,
which is also present
in using the CMB models with $N_{MB}= 8, 6, 3$ 
in Refs.~\cite{cw90},~\cite{pitt-anl},~\cite{zegrab-2}, respectively.
We also note that the similar many-parameters problem is also a concern
 in all approaches
of resonance extraction which require high precision fits 
of $\pi N$  data.
This common problem poses difficulties  
in assigning the errors for the determined
model parameters. We thus follow all previous works and will
only assign errors in
the determined $P_{11}$ resonance pole parameters which are determined 
non-linearly by the model parameters associated with meson-exchange
interactions $v_{MB,M'B'}$ and bare $N^*$ states.

Our fitting procedure is as follows.
We first adjust the parameters
associated with the meson-exchange 
interaction $v_{MB,M'B'}$ to
fit $P_{11}$ amplitude at low energies $W \lesssim 1.2$ GeV.
To control the number of parameters associated with bare $N^*$ states,
we then include only one bare $N^*$ state and try to fit the data in the 
entire considered energy region by adjusting its bare mass $m^0_{N^*}$ and
vertex function parameters $g_{MB}$ and $\Lambda_{MB}$. 
If this fails, we then also allow the parameters
associated $v_{MB,M'B'}$ to vary. If this fails again, we then
include one more bare $N^*$ state and repeat the process.
In the region below $W = 2$ GeV, we find that the considered
$P_{11}$ amplitudes  can be fitted 
with  one or two bare $N^*$ states. Most of the resulting cutoff parameters
are in the range of [$500$ MeV - $1500$ MeV], which are similar
to those in typical meson exchange
models~\cite{afnan,gross,sl96,ntuanl,juelich-1}. The bare $N^*$ masses are
searched within the range $m^0_{N^*} \leq 2500$ MeV. The
interpretations of  these  resulting bare $N^*$ parameters with
hadron structure calculations remain
to be developed. At the present time, they should be considered
 purely phenomenologically and only the extracted
resonance pole parameters have  well defined physical meaning.

Once a fit is obtained, we then apply the
 method of analytic continuation of Ref.~\cite{ssl09} to find
resonance poles, as also briefly described at the end of 
Sec.~\ref{sec2a}. 
The errors of the resonance parameters are then estimated by
using all values obtained in all fits we have performed.

For each of fits to be presented below, we assess its quality by evaluating
its $\chi^2$ per data point defined by
\begin{eqnarray}
\chi^2_{pd} &=& \sum_{i = 1,N_W}\frac{1}{N_{\text{data}}}
\left\{
\frac{\left|\text{Re}[T^{\text{model}}(W_i)] - \text{Re}[T^{\text{data}}(W_i)]\right|^2}
{\left|\text{Re}(\delta[T^{\text{data}}(W_i)])\right|^2} 
 + 
\frac{\left|\text{Im}[T^{\text{model}}(W_i)] - \text{Im}[T^{\text{data}}(W_i)]\right|^2}
{\left|\text{Im}(\delta[T^{\text{data}}(W_i)])\right|^2}\right\} \nonumber \\
\label{eq:chi2}
\end{eqnarray}
where $T^{\text{data}}(W_i)$ and $\delta[T^{\text{data}}(W_i)]$ 
are the values and errors
of the considered data, respectively;
$N_W$ is the number of the energy points where the data exist;
$N_{\text{data}} = 2 N_W$ is the number of the data points
(note that there are real and imaginary components at each energy point). 
We use the single energy solution SAID-SES as data in our fits, except in one fit
using CMB data (see below).
As a reference we
take the energy dependent solution SAID-EDS as $T^{\text{model}}(E)$ to
get $\chi^2_{pd}=2.94$, as
listed in the first row of Table I along with their values of 
$P_{11}$ resonance pole positions. 
Note that their sheet assignments are different
from ours since they do not have $\sigma N$ channel. Also 
they do not have a pole at higher energy region.

We now proceed to present our results by first recalling
the three $P_{11}$ poles 
extracted~\cite{sjklms10} from using the JLMS parameters.
They 
 are listed in the second row of 
Table~\ref{tab:p11-tab1} and the corresponding amplitudes
(solid curves) are compared with 
the SAID-EDS~\cite{said-1} (open circles) in Fig.~\ref{fig:p11-fig1}.
Here we note that the $\chi^2_{pd}$ from the JLMS fit listed in 
Table~\ref{tab:p11-tab1} is comparable to that of SAID-EDS.
In general, we find it is rather difficult to get a fit with 
$\chi^2_{pd} \leq 2.5$ within meson-exchange model,
mainly because the errors 
of SAID-SES are very small in $W \leq 1.45$ GeV region within which
the reproduction of the rapid sign changes of empirical
amplitude is rather
difficult due to the need of delicate balance
between the attractive and repulsive effects in different energy region.
 
In the following subsections, we present results from various fits by
varying the dynamical content of the EBAC-DCC model as described above
and using a model with a bare nucleon described in 
Sec.~\ref{sec2b}.

\subsection{1$N^*$-3p-H and 1$N^*$-3p-L fits}
\label{sec3a}

\begin{figure}[t]
\begin{center}
 \includegraphics[clip,width=0.8\textwidth]{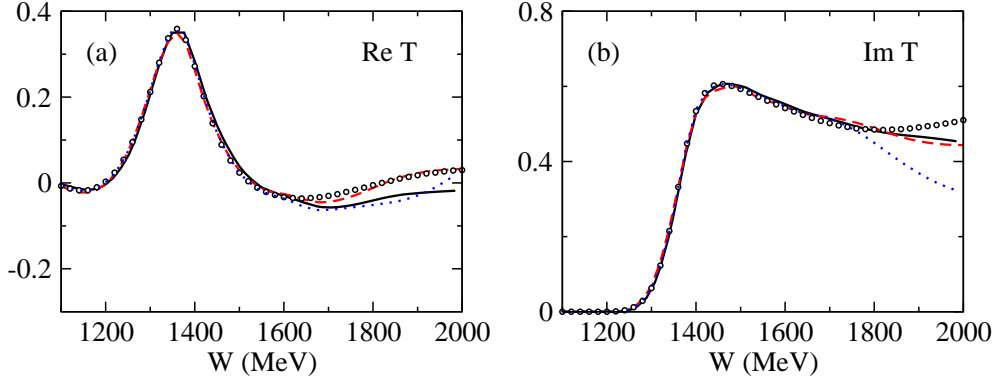}
\caption{\label{fig_said_1nstar}
(Color online)
The real (left) and imaginary (right) parts of the on-shell $P_{11}$
amplitudes as a function of the $\pi N$ invariant mass $W$ (MeV).
The solid curves are from the JLMS fit; the dashed (dotted) curves are 
from the 1$N^*$-3p-H (1$N^*$-3p-L) fit to the SAID-EDS~\cite{said-1} 
up to $W=2$ GeV ($W=1.6$ GeV); the open circles are the SAID-EDS~\cite{said-1}.
$T$ is unitless in the convention of Ref.~\cite{said-1}.
 \label{fig:p11-fig1} }
\end{center}
\end{figure}

We first consider the simplest variation of the JLMS fit by including only one
bare $N^*$ state, instead of two, to fit the SAID-SES solution.
In these  fits,  the parameters of meson-baryon interactions $v_{MB,M'B'}$
of Eq.~(\ref{eq:cc-mbmb}) are taken from JLMS. 
We also examine how the extracted
resonance poles depend on the data included in the fits.
Here we present results from two fits. The solution 1$N^*$-3p-H fits the
SAID-SES up to $2$ GeV, while the 1$N^*$-3p-L to only $1.6$ GeV.
These two fits are compared with the JLMS results in Fig.~\ref{fig:p11-fig1}.
The resulting resonance poles are listed in the
third and fourth rows of Table~\ref{tab:p11-tab1}.
We see that 
 the first two poles near the $\pi\Delta$ threshold ($\sim 1.3$ GeV) 
 are in good agreement with those from JLMS.
This suggests that these two poles are only sensitive to the data below 
about 1.5 GeV. The differences between these two fits and JLMS
at higher  $W >  $ 1.5 GeV mainly
affect the positions  of their third poles, as seen in Table~\ref{tab:p11-tab1}.

The results presented here also indicate 
that with the nonresonant amplitudes of JLMS,
only one bare N* state is sufficient to describe the $\pi N$ scattering data up
to 2 GeV. All of the fits to be presented below are obtained by
starting with  non-resonant amplitudes which are chosen to be different
from  that of JLMS by tuning the parameters of $v_{MB,M'B'}$.
It turns out that in these fits, using the procedure described above, 
two bare $N^*$ states are needed to
get comparable $\chi^2$.

\subsection{2$N^*$-3p and 2$N^*$-4p fits}
\label{sec3b}

Here we investigate the dependence of the extracted resonances
on the accuracy of the employed partial-wave amplitudes by considering
the SAID-SES solution which show some oscillating structure
in the  high $W \gtrsim 1.5$ GeV region. Such a structure is absent
in the SAID-EDS (open circles in Fig.~\ref{fig:p11-fig1}). 
From the empirical point of view, it raises the
question on whether the fits to the smooth SAID-EDS miss some resonance
physics of the original $\pi N$ data. Before more precise empirical amplitudes
are available, it is necessary to explore the extent to which these 
experimental uncertainties can affect the resonance extractions.
We explore this issue
by allowing the parameters
associated with meson-baryon interaction $v_{MB,M'B'}$ to deviate from the
JLMS values in varying these parameters
 along with the bare $N^*$ parameters in minimizing $\chi^2_{pd}$.
In general, the resulting
$\pi N\Delta$ and $\rho NN$ coupling constants from these new fits
are weaker than the JLMS values and hence give rather different
 non-resonant amplitudes
$t_{\pi N,\pi N}$.

\begin{figure}[t]
\begin{center}
 \includegraphics[clip,width=0.8\textwidth]{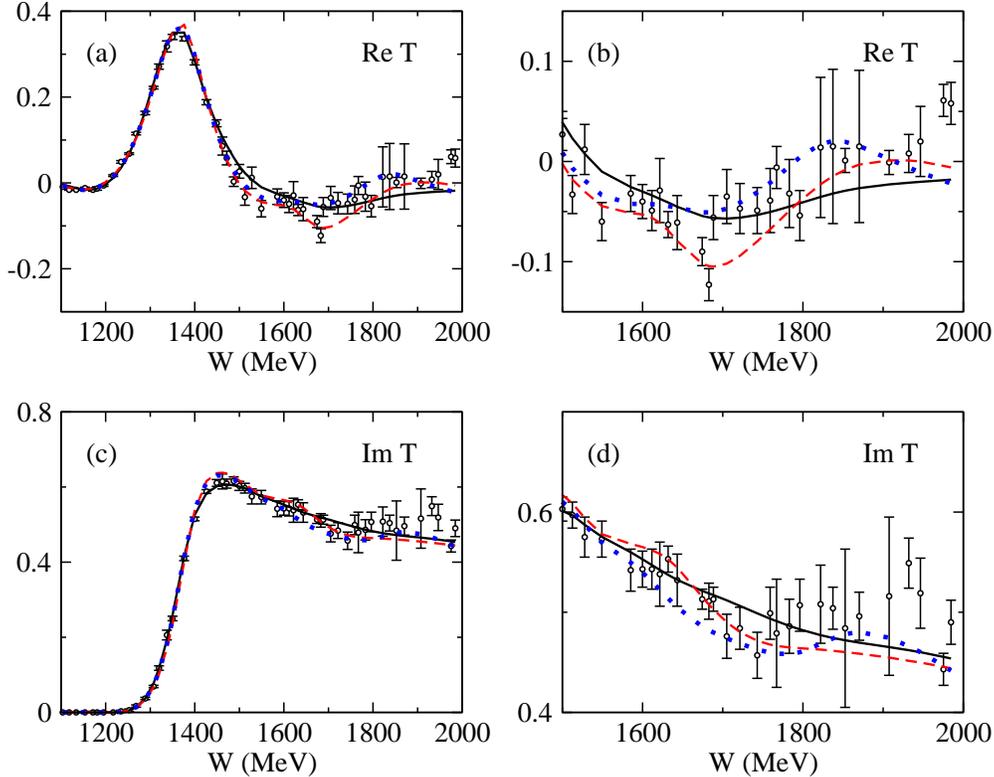}
\caption{\label{fig:fig5}
(Color online)
The real (upper panels) and imaginary (lower panels) parts of the  $P_{11}$
amplitudes as a function of the $\pi N$ invariant mass $W$ (MeV).
The JLMS (solid) results are compared with the results from the
2$N^*$-3p (dotted) and 2$N^*$-4p (dashed) fits.
The points with errors are from the SAID-SES~\cite{said-1}.
$T$ is unitless in the convention of Ref.~\cite{said-1}.
\label{fig:p11-fig2}
}
\end{center}
\end{figure}

We have obtained several fits which differ from each other
mainly in how the oscillating structure of the data at high $W$ are fitted.
The results from the 2$N^*$-3p (dotted curves) and 2$N^*$-4p (dashed curves)
fits are compared with the JLMS fit (solid curves) 
in Fig.~\ref{fig:p11-fig2}.
The resulting resonance poles
are listed in the 5th and 6th rows of Table~\ref{tab:p11-tab1}.
Here we see again the first two poles near the $\pi\Delta$ threshold from
both fits agree well with the JLMS fit. This seems to further support the
conjecture that these two poles are mainly sensitive to the
data  below $W\sim 1.5$ GeV where the SAID-SES has rather small errors.
However, the 2$N^*$-4p fit has one more pole at 
$M_R= 1630 -i45$ MeV. This
is perhaps related to its oscillating structure near $W\sim 1.6$ GeV
(dashed curves), as shown in the Figs.~\ref{fig:p11-fig2}(b) and~\ref{fig:p11-fig2}(d). 
On the other hand, this 
resonance pole could be
fictitious since the fit 2$N^*$-3p (dotted curve) with only three poles
are equally acceptable within the fluctuating experimental errors.
Our result suggests that it is important to have more accurate data
in the high $W$ region for a high precision resonance extraction.

\subsection{2$N^*$-4p-CMB fit}
\label{sec3c}

\begin{figure}[t]
\begin{center}
\includegraphics[clip,width=0.8\textwidth]{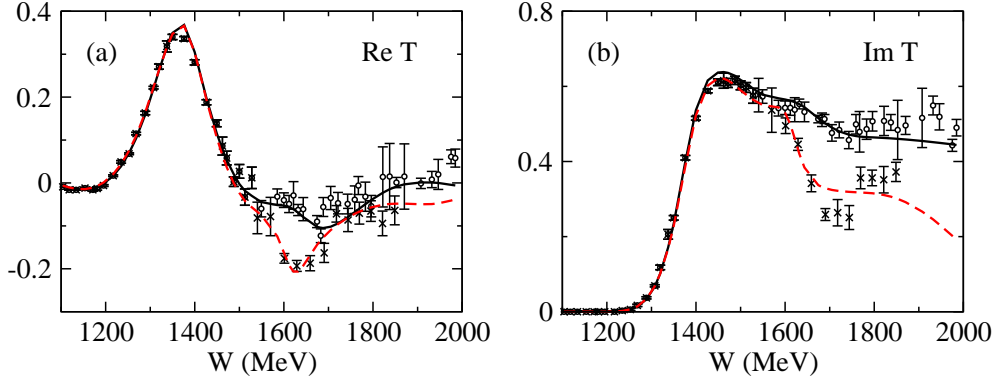}
\caption{\label{fig:p11-fig3}
(Color online)
The real (left) and imaginary (right) parts of the  $P_{11}$
amplitudes as a function of the $\pi N$ invariant mass $W$ (MeV).
The $2N^*$-4p fit (solid) and the $2N^*$-4p-CMB fit (dashed) are
compared with the data. 
The open circles with errors are from the SAID-SES~\cite{said-1},
and the crosses with errors are from the SAID-SES at $W<1.55$ GeV
and the CMB solution~\cite{cw90} at $W >$ 1.55 GeV. 
$T$ is unitless in the convention of Ref.~\cite{said-1}.
}
\end{center}
\end{figure}

To further explore the dependence of the resonance poles on
the data, we consider
a solution from CMB collaboration~\cite{cw90}. This solution differs
significantly from the SAID-SES mainly at $W > 1.55$ GeV. For our present 
purpose of investigating the stability of the lowest two poles near 
the $\pi\Delta$ threshold, we fit the data which is obtained from 
replacing SAID-SES in the high $W > 1.55$ GeV region by the CMB solution.
The results (dashed curves) from this fit
with all parameters allowed to vary within the EBAC-DCC model
are compared with
that of the 2$N^*$-4p (solid curves) in Fig.~\ref{fig:p11-fig3}.
 We see that both have oscillating behavior near $W \sim$ 1.6 GeV and
this could be the common reason why both have an addition pole 
near $W \sim 1.6$ GeV, as seen in rows 6 and 7 of Table~\ref{tab:p11-tab1}.
The large differences in their fits at high $W$ make their  poles near 
$W\sim$ 1.9 GeV very different; in particular their imaginary parts.
On the other hand, their lowest two poles near the $\pi\Delta$ threshold are
close to other fits discussed so far. 
This again supports the above observation that
these two poles are determined only by the data below $W < $ 1.5 GeV
which are reproduced very well in all fits.

\subsection{1$N_0$1$N^*$-3p}
\label{sec3d}

Here we consider the question concerning whether the analytic structure of
the employed reaction model in the $W \leq m_N + m_\pi$ unphysical 
region can strongly influence the resonance extractions.
We first note that most of the resonances listed by Particle Data
Group~\cite{pdg} (PDG) are from analysis which treat nucleon 
as a structureless basic degree
of freedom in describing the $\pi N$ reactions; such models are used in
SAID~\cite{said-1} and CMB~\cite{cw90}.
Similar simplification is used in formulating the EBAC-DCC model~\cite{msl07}.
On the other hand, a more elaborated approach has been taken to analyze $\pi N$
data using models within which the nucleon is made of 
a bare nucleon $N_0$ and meson clouds. Such 
models~\cite{afnan,gross,ntuanl,juelich}  
need to account for the nucleon pole condition, as described 
in Sec.~\ref{sec2b},
in fitting the $\pi N$ reaction data.
While all of these models give similar $P_{11}$
amplitudes from threshold $W_{th}=m_N+m_\pi$ to about 1.6 GeV, their analytic
structure as function of the complex energy could be very different in the
$W \leq m_N + m_\pi$ region where all dynamical 
models~\cite{afnan,gross,sl96,ntuanl,juelich,jlms07} have various
singularities due to the parameterization of the considered meson-baryon
interactions. This is discussed in Ref.~\cite{jklmss09}.
The question is whether such differences 
can lead to very different resonance poles.

\begin{figure}[t]
\begin{center}
 \includegraphics[clip,width=0.8\textwidth]{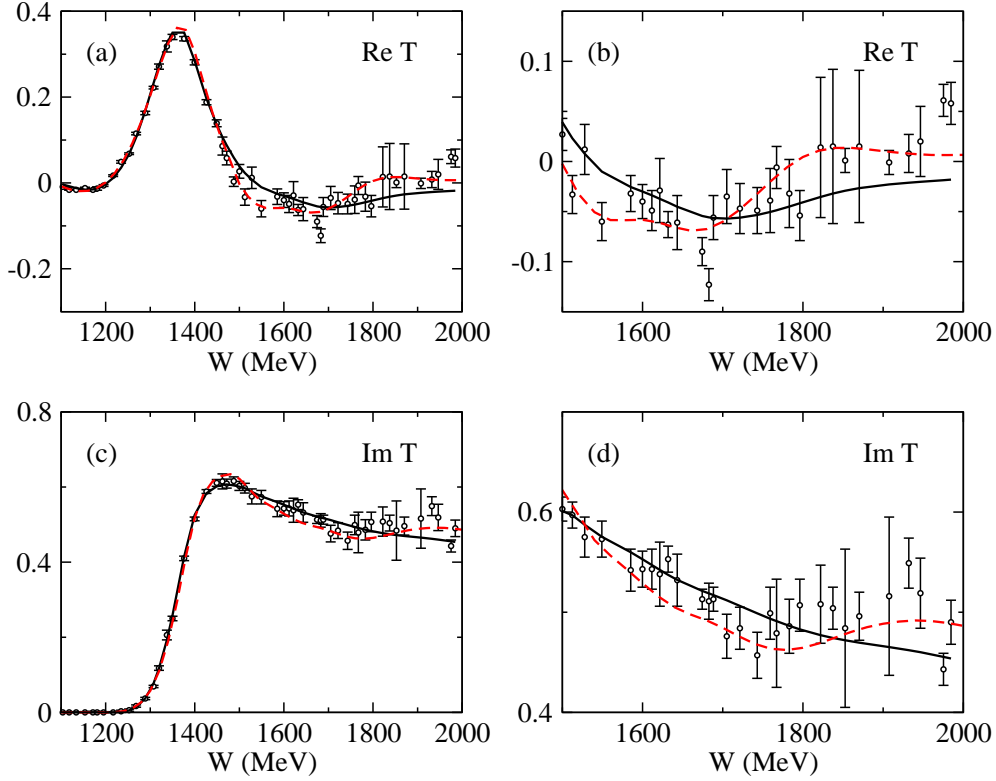}
\caption{\label{fig:p11-fig4}
(Color online)
The real (upper panels) and imaginary (lower panels) parts of the  $P_{11}$
amplitudes as a function of the $\pi N$ invariant mass $W$ (MeV).
The JLMS fit (solid) and
the 1$N_0$1$N^*$-3p fit (dashed) are compared with
the SAID-SES~\cite{said-1}. $T$ is unitless in the convention of Ref.~\cite{said-1}.}
\end{center}
\end{figure}

We investigate this issue by comparing the results presented above with
that from the fits 
using the model with a bare $N_0$ described in Sec.~\ref{sec2b}.
In these fits, the parameters of 
the meson-baryon interaction $v_{MB,M'B'}$ are adjusted along with the
parameters associated with $N_0$ and $N^*$ in fitting
the SAID-SES  up to $W=2 $ GeV under the nucleon pole 
condition~(\ref{eq:pole-m}) and~(\ref{eq:pinn-ff}).
The results from one of the fits (dashed curves)  are compared with 
the JLMS fits (solid curves) in Fig.~\ref{fig:p11-fig4}.
We see that the two fits agree very well below $W=1.5$ GeV, while
their differences are significant in the high $W$ region, as seen in the
right-hand-side panels of  Fig.~\ref{fig:p11-fig4}.
The resulting resonance poles are given in the  last row of 
Table~\ref{tab:p11-tab1}. Similar to all of the cases discussed above,
we also see here that the first two
poles near the $\pi\Delta$ threshold are close to those of JLMS.
Our results seem to indicate that these two poles are rather insensitive to
the analytic structure of the amplitude in the region below $\pi N$ threshold, 
and are mainly determined by the data in the region 
$ m_N+m_\pi\leq W \leq 1.6 $ GeV.
The third pole from this fit is close to that of JLMS, except that its
imaginary part is smaller as seen in the first and last rows of 
Table~\ref{tab:p11-tab1}.

\subsection{Averaged values of the extracted $P_{11}$ resonances}
To get the averaged values of the extracted $P_{11}$ resonance poles, we 
take the values listed in Table I except those from models 1$N^*$-3p-L
and 2$N^*$-4p-CMB which are obtained from fitting different sets of data,
as described above. We further omit the values 
from 2$N^*$-4p in the evaluation since it
has one more pole due to its oscillating behavior (dashed curves in Fig.2)
which needs further investigations although it is within the experimental
uncertainties. Our  values  
are listed in Table~\ref{tab:p11-tab3}. The errors are assigned by the 
differences between the largest and smallest values listed in Table I.

The model parameters from our fits are not relevant to the discussions 
given above
and are therefore not presented. These information is available upon 
requests.

\begin{table}[t]
\caption{\label{tab:p11-tab3} Averaged values of the extracted $P_{11}$
resonances [listed as ($\text{Re}M_R$, $-\text{Im} M_R$) in the unit of MeV].
Here we identify these poles with the states listed by PDG~\cite{pdg}.
The location of the pole is specified by, e.g.,
$(s_{\pi N},s_{\eta N},s_{\pi\pi N},s_{\pi\Delta},s_{\rho N},s_{\sigma N})=(upuupp)$,
where $p$ and $u$ denote the physical and unphysical sheets for a
given reaction channel, respectively.
} 
\begin{ruledtabular}
\begin{tabular}{ccc}
States        & Location  &     Averaged values (MeV) \\ \hline
$N(1440)$     & $(upuupp)$  &$(1363^{+9}_{-6}  , 79^{+3}_{-5})$ \\
              & $(upuppp)$  &$(1373^{+12}_{-10}, 114^{+14}_{-9})$ \\
\hline
$N(1710)$     & $(uuuuup)$&$(1829^{+131}_{-65}, 192^{+88}_{-110})$ \\
\end{tabular}
\end{ruledtabular}
\end{table}
\section{Summary and Discussions}
\label{sec4}

In this work we have investigated the extraction of $P_{11}$ nucleon resonances.
By performing extensive fits to SAID-SES,
we show that two resonance poles near the $\pi \Delta$ threshold
are stable against large variations of parameters of  meson-exchange mechanisms
 within EBAC-DCC model~\cite{msl07}.  This two-pole structure is also obtained 
in an analysis based on a model with a bare nucleon state.
Our results indicate that
the extraction of $P_{11}$ resonances is insensitive to the analytic structure
of the amplitude in the region below $\pi N$ threshold.

By performing different fits to the structure of
SAID-SES as well as the old, perhaps also outdated, CMB data, we
demonstrated that
the number of  poles in the
1.5 GeV $\leq W \leq $ 2 GeV region could be more than one.
Thus our determination of the resonance poles in this
higher $W$ region is not so conclusive. We can only report
one pole near $N(1710)$ state listed by PDG, in agreement with several
previous analyses.
Our results  indicate the need of more accurate $\pi N$ reaction data in
the $W > $ 1.5 GeV region for high precision resonance extractions.
In particular, accurate inelastic amplitudes for $\eta N$, $\pi \Delta$,
$\rho N$, and $\sigma N$ channels are highly desirable for our 5-channels
analysis.
This will allow  simultaneous fits to both elastic and inelastic
 amplitudes to firmly determine
the nucleon resonances in the
1.5 GeV $\leq W \leq $ 2 GeV region. The importance of performing multi-channel
fits was demonstrated  in a recent
3-channels CMB analysis~\cite{zegrab-2} in which it was shown that 
a simultaneous fit to both $\pi N \rightarrow \pi N$ and
$\pi N \rightarrow \eta N$ is needed to establish $N(1710)$ state.
Thus it is important to obtain more extensive data of
$\pi N$ reactions including polarization
observables such that high precision
partial-wave amplitude analysis of $\pi N \rightarrow \pi\pi N$
data can be performed. Such experiments are possible in new hadron facility
J-PARC in Japan. 

Finally,  we like to  mention that the analysis of electromagnetic 
$\pi$ and $2\pi$ production data can help confirm the nucleon resonances 
extracted from $\pi N$ reaction data, although its main objective is to 
extract electromagnetic properties of nucleon resonances. 
On the other hand, some resonances, which have small branching ratios
to $\pi$ and $2\pi$ channels and have large ones for $KY$ and $\omega N$
channels, could  be identified 
from analyzing the data
of $\gamma N \rightarrow KY, \omega N$ which have been accumulated 
extensively in recent
years. This is also an important task in $N^*$ study  
before the hadronic data for these channels
become extensive at new hadron facility.

\begin{acknowledgments}
This work is supported 
by the U.S. Department of Energy, Office of Nuclear Physics Division, under
Contract No. DE-AC02-06CH11357, and Contract No. DE-AC05-06OR23177
under which Jefferson Science Associates operates Jefferson Lab, and by
the Japan Society for the Promotion of Science,
Grant-in-Aid for Scientific Research(C) 20540270. 
This research used resources of the National Energy Research Scientific Computing Center, which is supported by the Office of Science of the U.S. Department of Energy under Contract No. DE-AC02-05CH11231. 
\end{acknowledgments}

\end{document}